# Exploration and prediction of topological electronic materials based on first-principles calculations

Hongming Weng[1,2], Xi Dai[1,2], and Zhong Fang[1,2]

[1]*Beijing National Laboratory for Condensed Matter Physics, and Institute of Physics, Chinese Academy of Sciences, Beijing 100190, China*

[2]*Collaborative Innovation Center of Quantum Matter, Beijing, China*

The class of topological insulator materials is one of the frontier topics of condensed matter physics. The great success of this field is due to the conceptual breakthroughs in theories for topological electronic states and is strongly motivated by the rich variety of material realizations, thus making the theories testable, the experiments operable, and the applications possible. First-principles calculations have demonstrated unprecedented predictive power for material selection and design. In this article, we review recent progress in this field with a focus on the role of first-principles calculations. In particular, we introduce the Wilson loop method for the determination of topological invariants and discuss the band inversion mechanism for the selection of topological materials. Recent progress in quantum anomalous Hall insulators, large-gap quantum spin Hall insulators, and correlated topological insulators are also covered.

## Introduction

In recent years, we have witnessed rapid progress in the field of topological electronic materials, particularly topological insulators (TIs)—a new state of quantum matter with great potential for applications in future and emerging electronic devices and technologies. Topology, a word mostly used in mathematics, is now frequently used to describe and classify the electronic structure of materials. "Topological electronic materials" refers to materials whose electronic structures have nontrivial topology in momentum space; examples include TIs,[1-3] topological semimetals,[4–11] and topological superconductors.[2,12-15] One of the most important characteristics of topology is its robustness against local deformations, or, in the language of physics, its insensitivity to environmental perturbations, thus making topological electronic materials promising for future applications.

The idea of topological electronic states can be traced back to the discovery and systematic studies of the quantum Hall effect (QHE) in the 1980s,[16,17] which suggested that symmetry alone cannot describe all possible phases of condensed matters. New quantum numbers, namely topological numbers, are required to go beyond Landau symmetry-breaking theory which points out that the state of matter can be described by a local order parameter corresponding to the symmetry broken during the phase transition. Different quantum Hall states all have the same symmetry, but they are different by a topological invariant, called the Chern number[18] or TKNN (Thouless-Kohmoto-Nightingale-den Nijs) number[19], which is usually labeled by an integer $Z$. It physically corresponds to the number of edge states or the winding number of the electron wave function in the two-dimensional (2D) Brillouin zone (BZ).[20]

For a long time, however, this conceptual breakthrough did not have much relevance to materials science, because the QHE is due to Landau level formation in 2D electron gases under strong external magnetic field. Under such extreme conditions, details of the material electronic band structure become irrelevant to the physics of QHE. In this sense, the lattice model for QHE proposed by Haldane in 1988 is very stimulating.[21] His results suggested that certain materials can have topologically non-trivial electronic band structure characterized by a non-zero Chern number (and called Chern insulator). They can support a similar QHE effect even without an external magnetic field (and of course no Landau levels) which is referred to as the quantum anomalous Hall effect (QAHE). Thus, topology of electronic band structure was related to the realistic materials, and, as a result, can benefit from the rich variety of material choices.

Haldane's idea was not realized until recent years. The QAHE was theoretically predicted in magnetically doped TIs, such as in a Mn-doped HgTe quantum well[22] and Cr-doped $Bi_2Te_3$, $Bi_2Se_3$, and $Sb_2Te_3$ family TIs,[23] and experimentally observed (for the latter).[24] Before this, however, the real step toward topological electronic materials was mostly due to the discovery of TIs in system keeping time reversal invariant. Kane et al.[25,26] and Bernevig et al.[27] proposed that a new topologically non-trivial electronic state can be realized in materials protected by time-reversal symmetry and can support the quantum spin Hall effect (QSHE). The topological invariant to characterize this non-trivial state is $Z_2$.[26] Bernevig et al.[28] further predicted that a HgTe/CdTe quantum well structure is a candidate for such a state; this theoretical prediction was experimentally confirmed by König et al. in 2007.[29]

Different from the Chern insulator, which is defined only for a 2D system, the idea of time-reversal invariant TIs can be naturally extended from 2D to 3D,[30-32] and up until now, most of the discovered TI materials have in fact been 3D compounds.[33] The $Bi_{1-x}Sb_x$ alloy was first predicted[30,34] and observed[35] to have topologically nontrivial conducting surface states, though its surface states are complex. The standard examples of 3D TIs[36] are the $Bi_2Te_3$, $Bi_2Se_3$ and $Sb_2Te_3$ families of compounds with single Dirac cone on the surface. These were also theoretically predicted[37] and experimentally observed.[38,39] More recently, the field of topological electronic materials has been extended from Chern insulators and $Z_2$ TIs to topological metals/semimetals, which need a Fermi surface Chern number to characterize the topological properties.[10,40] The theoretically predicted compounds $Na_3Bi$[10] and $Cd_3As_2$[11] were experimentally confirmed recently.[41-44]

Theoretical predictions, particularly first-principles electronic structure calculations based on the density functional theory (DFT),[45] have played an important role in the exploration of almost all topological electronic materials. This is not a coincidence, but rather due to the following reasons: (1) to describe complicated electronic band structures of real materials, particularly with spin–orbit coupling (SOC), the first-principles calculations are needed; (2) first-principles calculations nowadays are accurate up to 90% in predicting various physical properties for many "simple" materials (i.e., weakly correlated electronic materials); (3) topological electronic properties of materials are robust, non-perturbative, and not sensitive to small error bars. Despite these advantages of first-principles calculations and their great success in this field, the numerical determination of topological invariants (such as integer $Z$ or $Z_2$ numbers) is still technically challenging, because those numbers are related to the phases of eigen wave functions, which are gauge dependent and randomized in most of the calculations. In addition, we also note that: (1) it is well known that band gap is underestimated in either the local density approximation (LDA) or the generalized gradient approximation (GGA) of exchange-correlation potential; (2) present first-principles calculations based on LDA or GGA cannot treat strongly correlated systems properly; (3) evaluation of the Berry phase[46,47] (the quantum geometric phase accquired by eigenstates after a adiabatic loop change of hamiltonian) and topological numbers may require some additional complexities, such as fine $k$-points meshes, the gauge-fix condition, etc.

In this article, we start from a simple introduction to the topological properties of electronic states in momentum space, avoiding the use of complicated mathematic formulations, and present a simplified view. In particular, we discuss some of the computational details for the determination of topological numbers by the Wilson loop method, and the corresponding boundary state calculations. From the viewpoint of materials science, we discuss some general guidelines, such as the band inversion mechanism, for materials selection and design. We focus on some recent progress, including the Chern insulator, large-gap 2D TIs, and electron-electron correlated TIs. In the last part, we discuss some remaining issues, problems, and perspectives for the future.

**Twisted bands and topological invariants**

In an insulator, the energy gap at the Fermi level separates the electronic states into occupied bands and unoccupied bands. At zero temperature, there are no states at the Fermi level, and therefore no electronic conductivity. From this point of view, all insulators should behavior in the same way. However, Chern insulators and topological insulators inform us that insulators can be further classified into trivial and non-trivial ones according to their topological properties of occupied electronic states in momentum space. Topologically non-trivial insulators have twisted bands and can be characterized by the non-zero topological invariants, such as $Z$ or $Z_2$. Insulators with infinitely large size (i.e., without a boundary) still look the same; however, in the presence of a boundary (say to vacuum), the topologically non-trivial insulators always support gapless boundary states protected by their topological properties, as distinguished from the trivial insulators. The existence of gapless (conducting) boundary states is responsible for the dissipationless quantum transport properties, such as the QAHE[23,24] and the QSHE,[28,29] which are useful for emerging technologies with ultralow power consumption. There are several in-depth review articles on topological insulators[1,2] in which readers can learn the fundamental mathematics and physics associated with the topological nature of insulators. See the Introductory article in this issue.

In a periodic solid crystal, the topology of electron bands and its twist in momentum space are due to the non-vanishing Berry phase[46] and the gauge field,[47] and can be also related to the Wannier functions[48,49] (a set of locallized orbitals for representation of Bloch wavefunction in solid) and their centers.[2,50] Let us consider a 2D system with periodic boundary condition, whose occupied eigen wave functions are given as the Bloch states $\psi_{nk}(r) = e^{ik \cdot r} u_{nk}(r) e^{i\phi_n(k)}$ with $n$ the band index and $u_{nk}(r)$ the cell-periodic part. We have explicitly written out the arbitrary phase $\phi_n(k)$, which is not assigned by the Schrödinger equation and depends on the explicit gauge choice in each calculation. The momentum $k = (k_x, k_y)$ is defined in the 2D BZ, which is a torus (a closed manifold without boundary) due to the periodic boundary condition. Considering the Berry connection $A_n(k) = i \langle u_{nk} | \nabla_k | u_{nk} \rangle$ (which is gauge dependent) and the Berry curvature $\Omega_n(k) = \nabla_k \times A_n(k)$ (which is gauge independent), the following equation holds as claimed by the Chern theorem:[18]

$$\sum_n \int\!\!\int_{BZ} \Omega_n(k) \cdot dS = 2\pi Z . \tag{1}$$

Namely, the surface integration of Berry curvature over a torus must be an integer multiple of $2\pi$. The integer number $Z$ is a topological invariant, the Chern number. An insulator with a non-zero $Z$ number is a Chern insulator, which supports chiral edge states[21] and the QAHE.[23,24] In such a case, the *Z* number is physically meaningful, and is the number of chiral edge states, similar to that in the QHE.[17,20]

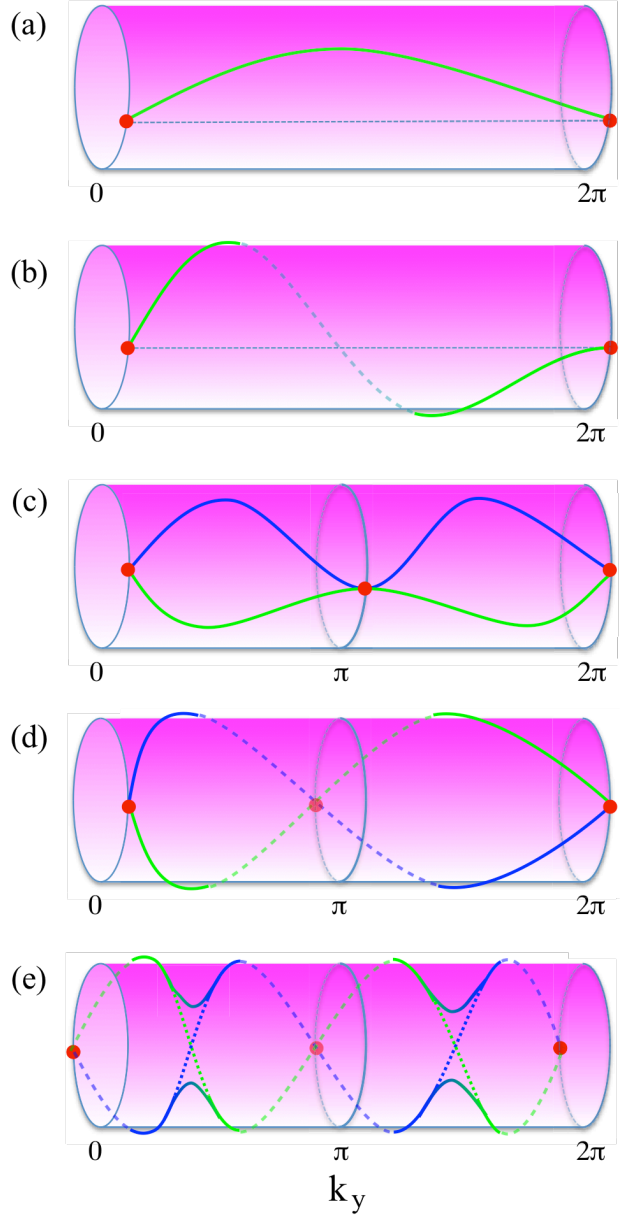

**Figure 1**. Schematic illustration of band twist in 2D Brillouin zone. The evolution of θ($k_y$) is plotted over a cylinder surface, whose azimuth is θ and longitudinal axis is $k_y$. (a) Trivial insulator with Z = 0; (b) non-trivial Chern insulator with *Z* = 1; (c) trivial 2D insulator with $Z_2$ = 0; (d) non-trivial 2D TI with $Z_2$ = 1; and (e) trivial 2D insulator with |*Z*| = 2, but $Z_2$ = 0 (see text for detailed explanation.) These figures are re-plotted based on Reference 52.

The band topology discussed previously can be understood intuitively as a band twist. Here we consider the simplest case: a 2D lattice with only one single occupied band (the band index $n$ can therefore be neglected). Let us chose a special gauge that is smooth and periodic along the $k_x$ direction, but not necessarily along the $k_y$ direction (this choice is always possible[51]). Thus, we can do the integration in Equation 1, explicitly for the $k_x$ direction with fixed $k_y$ and obtain

$$2\pi Z = -\int_0^{2\pi}\int_0^{2\pi} dk_x dk_y (\partial_x A_y - \partial_y A_x)$$
$$= \int_0^{2\pi} dk_y \partial_y (\int_0^{2\pi} dk_x A_x(k_x, k_y))$$
$$= \int_0^{2\pi} d\theta(k_y). \qquad (2)$$

Here, $\theta(k_y) = \int_0^{2\pi} dk_x A_x$ is an angle calculated from the 1D integration of $A_x(k_x, k_y)$ along the $k_x$ axis for each fixed $k_y$. We can plot θ ($k_y$) over a cylinder surface (in the cylinder coordinates), as shown in **Figure 1**. $k_y$ is plotted along the longitudinal direction, and the azimuth is the angle θ ($k_y$) for each fixed $k_y$. Moving from $k_y = 0$ to $k_y = 2\pi$, we can see the difference between the trivial insulator ($Z = 0$) and the Chern insulator ($Z \neq 0$). The winding number of θ ($k_y$) over the cylinder surface is zero for the former (Figure 1a), and non-zero for the latter (Figure 1b). In this way, the Chern number can be related to the winding number of eigen wave functions, and see that the Chern insulator has "twisted" energy bands.

The above is for Chern insulators, which break the time reversal symmetry (TRS). In the presence of TRS, however, the situation is different, because the total $Z$ number (after summation over the band index $n$) should always be zero due to the Kramers degeneracy, the degeneracy of spin up and down eigenstates in time reversal invariant system. In this case, the total number of occupied electronic states must be even, and we can, in principle, divide them into two classes, $\psi^I$ and $\psi^{II}$, which are TRS related and are called Kramers pairs. Evaluating the Chern number for each class independently, if one class has Chern number $Z$, the other one must have Chern number $-Z$ (due to the TRS), and the total Chern number of the system is zero. The question is whether $|Z|$ can be used as the definition of the topological properties. The answer is no. This is because there are many different ways to divide the occupied wave functions into two classes in the presence of band degeneracies. A different way of dividing may lead to different $|Z|$ numbers, which are different by an even number.[32,50] Therefore, although $|Z|$ is not a well-defined number, $|Z|$ mod 2 is a topological invariant, which is called the $Z_2$ number, and can be used for the definition of topological insulators with TRS.[26,30-32]

Let us again consider the simplest example of a 2D insulator with only two occupied eigenstates, $\psi_{1k}$ and $\psi_{2k}$, which are related by TRS. The two states form a Kramers pair and must be degenerate at the time-reversal-invariant momentum (TRIM) of the BZ. We can evaluate the Berry connection and check the winding number of each occupied eigenstate (as we have done above for the Chern insulator). In general, there are three situations (as shown in Figure 1c–e). First, if the winding numbers of two states are zero (i.e., $|Z| = 0$), this is the trivial situation with $Z_2$ index $v = 0$ (Figure 1c). Second, if the winding numbers of two states are ±1 (i.e., $|Z| = 1$), this is the non-trivial topological insulator with a $Z_2$ index $v = 1$, where the crossing of two $\theta(k_y)$ curves at $k_y = \pi$ is protected by TRS (Figure 1d). Third, if the winding numbers of two states are ±2 (i.e., $|Z| = 2$), the two $\theta(k_y)$ curves must cross at some $k_y$ other than the TRIM. Such crossings are not protected by TRS, and can be removed by small perturbations, which drive the crossings into anti-crossings. As a result, the original winding number $|Z| = 2$ is reduced to zero, and the system becomes equivalent to the trivial case with a $Z_2$ index $v = 0$ (Figure 1e).

The 2D $Z_2$ topological insulator with TRS is different from a Chern insulator in the sense that it supports helical (rather than chiral) edge states and the QSHE.[25-28] Here, the helicity of edge states means the orientation of its spin and momentum is locked, and the scattering between counter-propagating channels is forbidden within TRS. The $Z_2$ number can be physically related to the number of helical edge state pairs mod 2. The Chern number is defined only for 2D insulators, while the idea of the $Z_2$ number can be extended from 2D to 3D compounds[30-32] by using the 2D $Z_2$ numbers as building blocks. For 3D insulators, there are four independent indices for $Z_2$ numbers, $v_0(v_x v_y v_z)$, as discussed in Reference 30. Here $v_0$ is for the total $Z_2$ number, and $v_{x,y,z}$ are for the $Z_2$ numbers in the $x$, $y$, $z$ directions, respectively. Using our scheme presented prievously for the 2D $Z_2$ number, we can obtain the three weak $Z_2$ indices $v_{x,y,z}$ directly from the three 2D $Z_2$ numbers evaluated for the $k_x = \pi$, $k_y = \pi$, and $k_z = \pi$ planes respectively. For the total $Z_2$ index $v_0$, however, we have to consider two planes together. We can choose the $k_x = 0$ and $\pi$ planes, or $k_y = 0$ and $\pi$ planes, or $k_z = 0$ and $\pi$ planes. If the $Z_2$ numbers of two planes are different, we have $v_0 = 1$, otherwise, if they are the same, we have $v_0 = 0$.

**Wilson loop method for the evaluation of topological invariants**

In real materials, the band structures are usually very complicated, and the determination of topological numbers becomes essential and computationally demanding. The band degeneracy, either accidentally or due to symmetry, makes the numerical determination of phases of wave functions a tough task. Here the Wilson loop method is presented for determining topological indices efficiently.[51-53] This method is computationally easy, and is equally applicable for Chern insulators, $Z_2$ topological insulators (2D and 3D), and the topological crystalline insulators,[54,55] which have attracted a lot of recent interest.

For band insulators with inversion symmetry, the $Z_2$ indices can be easily computed as the product of half of the parity (Kramers pairs have identical parities) numbers for all the occupied states at the TRIM points.[56] In such a case, parity is a good quantum number for the TRIM points and can be computed from the eigen wave functions obtained from first-principles calculations. Up until now, this method has been frequently used and is very efficient.

Unfortunately, for general cases where inversion symmetry is absent, evaluation of the $Z_2$ number becomes difficult. There are mainly three different ways. The *first method* computes the $Z_2$ numbers using the integration of Berry curvature $\Omega(k)$ over half of the BZ.[57] In order to do so, one has to set up a fine mesh in $k$-space and calculate the corresponding quantities for each $k$ point. Since the calculation involves the Berry connection $A(k)$, one has to numerically fix the gauge on the half BZ, which is not easy for the realistic wave functions obtained by first-principles calculations. *The second method* starts from an artificial system with inversion symmetry, and then smoothly deforms the Hamiltonian toward the realistic one without inversion symmetry. If the energy gap never closes at any point in the BZ during the deformation process, the realistic system must share the same topological nature with the initial reference system, whose $Z_2$ number can be easily counted by the parity eigenvalue formula.[56] Unfortunately, making sure that the energy gap remains open on the whole BZ is very difficult numerically, especially in 3D. The third method directly calculates the boundary (edge or surface) states. Due to open boundary conditions, the first-principles calculation for the boundary states is numerically heavy. The Wilson loop method that we present here has the following advantages: first, it uses only the periodic bulk system; second, it does not require any gauge-fixing condition—thereby greatly simplifying the calculation; and third, it can be easily applied to general systems with or without inversion symmetry.

As discussed in the previous section, to determine the topological number (either $Z$ or $Z_2$), we must determine the key quantity $\theta(k_y)$ angle and its 1D evolution. In the single band model, the $\theta(k_y)$ angle is simply given as the 1D integration of $A_x(k_x, k_y)$ along the $k_x$ axis. However, for real compounds with multiple bands, its computation needs the Wilson loop method.[52] Once the $\theta$ angle can be computed, all the topological numbers can be determined easily following the previous discussions. For the mathematical details of this method, refer to Reference 52.

The band-dependent phase $\theta_m(k_y)$ is exactly what we need to determine the winding number and the corresponding topological numbers, as discussed in the previous section. In practice, for a system with many bands, an arbitrary reference line can be drawn on the cylinder surface, parallel to the $k_y$ axis, and then a count taken of how many times the evolution lines of $\theta_m(k_y)$ cross the reference line.[52] For the calculation of $Z_2$ in time-reversal invariant systems, we can further reduce the computational task by considering only the $0 < k_y < \pi$ part. As an example, **Figure 2** shows the calculated $\theta_m(k_y)$ and their evolution for the 3D topological insulator $Bi_2Se_3$ in the $k_z = 0$ and $\pi$ planes.[52] The $\theta_m(k_y)$ lines cross the reference line an odd number of times for the $k_z = 0$ plane, but an even number of times for the $k_z = \pi$ plane. Therefore, it is topologically non-trivial with $Z_2 = 1$.

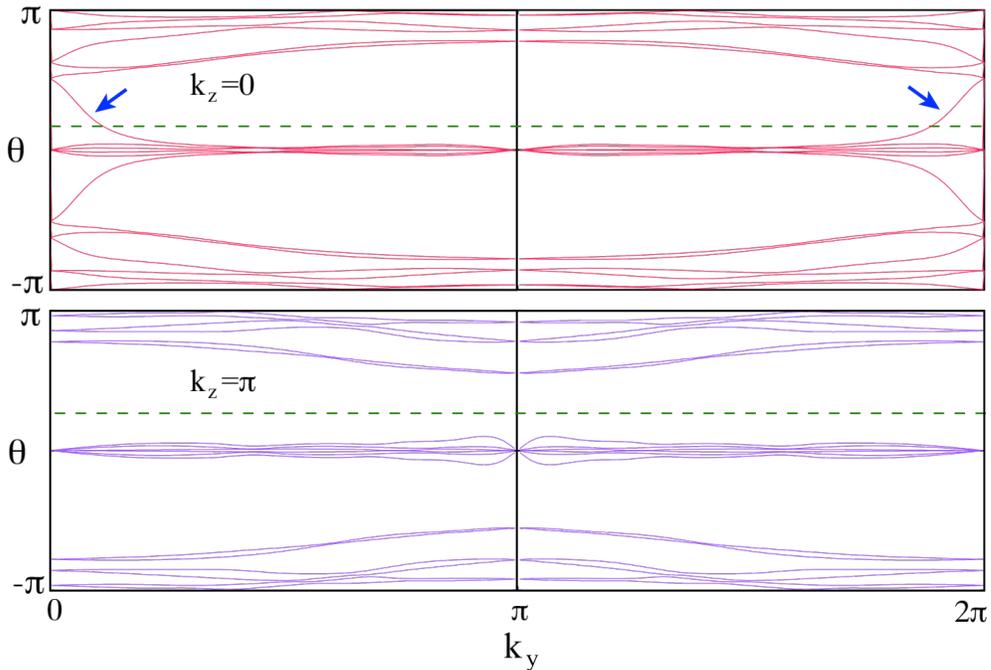

**Figure 2**. An example of the Wilson loop method for the 3D strong topological insulator $Bi_2Se_3$ with $Z_2 = 1$. The calculated $\theta_m$ ($k_y$) for all occupied bands ($m$ is the band index) are plotted as a function of $k_y$. The dashed lines are arbitrary reference lines. $\theta_m$ ($k_y$) crosses the reference line odd number of times in half of the $k_y$ direction ($k_y$ varies from 0 to $\pi$ or from $\pi$ to $2\pi$) for the $k_z = 0$ plane (upper panel), but even number of times for the $k_z = \pi$ plane (lower panel). These are re-calculated and re-plotted based on Reference 52.

Finally, in the whole process of the calculations, the advantages of Wannier functions can be taken. Particularly, the Maximally localized Wannier functions (MLWFs) has been widely used.[58,59] For example, after the self-consistent electronic structure calculations, the MLWF for the energy bands close to the Fermi level can be constructed, which can be reproduced usually to very high accuracy. In this process, however, some of the unoccupied bands need to be included.[51] Once the MLWF are obtained, calculations can be performed for the eigen wave functions and their inner products for any $k$-points using the interpolation technique.[60,61] This greatly simplifies the task for the computation of topological numbers. In addition, the MLWFs are also useful for the calculations of boundary states, as discussed later.

**Boundary state calculations**

After the determination of bulk topological numbers by either of the methods presented previously, the calculation of boundary states of real compounds becomes necessary. This is not just for the purpose of further confirmation of the topological properties, but is also very useful for direct comparison to experiments, such as using angle-resolved photoemission spectroscopy (ARPES) and scanning tunneling microscopy (STM).

In practical terms, two methods are usually used for the calculation of boundary states. First, the boundary states can be directly calculated from first-principles calculations by using a supercell of object material slab intercalated with vaccum layer, as is usually done for other surface or edge calculations.[62] In this way, the atomic details, such as the surface or edge terminations, the atomic relaxations or reconstructions, possible defects or absorptions, can be carefully treated and studied. This method is straightforward, although it is computationally demanding. Due to the presence of boundary states, we have to use a very large supercell with a thick vacuum region to avoid possible mutual coupling between the boundaries. A convergence test as a function of cell size is usually necessary.

Second, the advantage of MLWF can be taken and the boundary states calculated from the Green's functions (please see solid state physics textbook) of a semi-infinite system.[63] The MLWF can be used as a basis to construct an effective low-energy Hamiltonian $H_{MLWF}$, which can be regarded as a tight-binding model with its hopping parameters determined from first-principle electronic structure calculations. The maximally localized property of MLWF guarantees short range hopping. Using $H_{MLWF}$ as building blocks, a large Hamiltonian $H_{semi}$ can be constructed for the semi-infinite system with only one boundary. This $H_{semi}$ is a block tridiagonal matrix if only the hopping terms are considered between the nearest-neighbor building blocks. Then, the iterative method[64-66] can be used to solve the problem and get the projected Green's functions on to the boundary. From this boundary Green's functions, the necessary information can be obtained, such as the charge density of states and the spin density of states at the boundary.[34,67] This procedure is an approximate one, and it cannot treat the atomic details at the boundary precisely. However, it provides the most important information on the gapless boundary states of TIs.

**Band inversion mechanism for materials selection**

The most intuitive picture and practically useful guideline for searching for topological insulators is the band inversion mechanism. It involves the energy order switching of low energy electronic bands around certain TRIM points in the Brillouin zone. The mechanism was first demonstrated for the HgTe/CdTe quantum well structure[28] and was later found to be applicable to many other materials, including 2D and 3D TIs[33,37,68] and Chern insulators.[23]

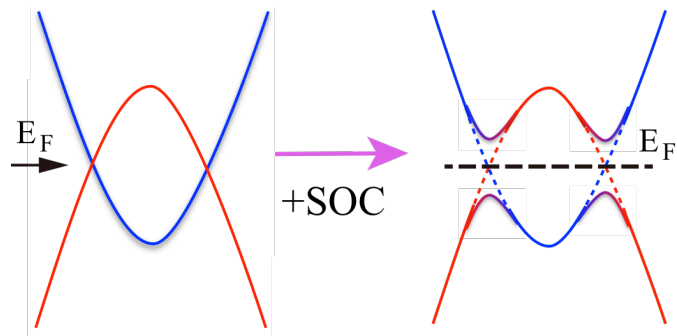

**Figure 3**. Schematic plot for the band inversion mechanism as discussed in the text. SOC means spin-orbit coupling.

As schematically illustrated in **Figure 3**, we consider two bands (indicated in red and blue, respectively) close to the Fermi level around the Γ point. Far away from the Γ point, the red state is occupied and energetically lower than the blue state. However, close to the Γ point, they have inverted energy order with the blue state lower and occupied at Γ. Assuming that the two bands do not couple when the SOC is absent, they must cross at certain *k*-points away from Γ. In the presence of the SOC, the two bands will, in general, couple together and open up a gap at the crossing point. As a result, the system becomes an insulator, and most importantly, this gapped state may be topologically non-trivial depending on the coupling term.

To be more explicit, we consider two parabolic bands with the simplest energy dispersion $E_\pm(k) = \pm(-m + k^2)$. The + and − signs are for the blue and red bands, respectively. The band inversion happens when $m > 0$ is satisfied. Using two bands as a basis, an effective 2×2 Hamiltonian is constructed:

$$H_{eff}(k) = \begin{bmatrix} E_+(k) & M(k)^* \\ M(k) & E_-(k) \end{bmatrix},$$ (3)

where $M(k)$ is the coupling between the two bands and determines the topological properties of the resulting electronic structure. If $m < 0$ (i.e., without band inversion), in general, nothing is expected but a normal insulator. However, if $m > 0$ for the band inverted system, we may find many interesting topological phases depending on the choice of $M(k)$. For example, if $M(k) = (k_x \pm i k_y)^n$ ($n \geq 1$), we will get a Chern insulator with Chern number $Z = n$ (for a 2D system). For time reversal invariant systems, the two bands should be replaced by four bands with each color (blue and red) representing a Kramer pair of bands, and the model Hamiltonian becomes 4×4. It is block diagonal in the simplest form, such as the Hamiltonian proposed for HgTe/CdTe quantum well.[28] If there is a form of $M(k) = (k_x \pm i k_y)^{2n-1}$, a TI with $Z_2 = 1$ is expected. In the presence of SOC, these forms of the coupling term $M(k)$ are in general allowed and depend on the material details. The band inversion may occur between the *s–p*, *p–p*, *p–d*, or *d–f* states, which leads to different material classes of TIs.[68]

The band inversion mechanism discussed previously is useful from a practical view-point, yet it is a crude picture for real compounds. First, this picture comes from the low-energy effective Hamiltonian of a continuous model. If the model is put onto a lattice, we have to be careful about the number of band inversions. Particularly for $Z_2$ TIs, if the band inversion happens an even number of times or it happens around an even number of TRIM points, we will return to trivial insulators. Second, real compounds may have many low-energy bands, which must be carefully analyzed. Third, the possible $M(k)$ terms can also be complicated and need careful treatment. In any case, the topology of the whole electronic structure must be determined by first-principles calculations, as presented in previous sections.

**Topological electronic materials**

First-principles calculations based on density functional theory have played an important role in studies on topological electronic materials. Many compounds have been theoretically predicted and experimentally confirmed. Although those computational techniques are still not perfect, they have demonstrated their power in materials selection and design, thus greatly motivating the field. It is a hard task to review and summarize the whole field and all known topological materials completely. Instead, here we discuss some recent interesting progress, in which first-principles calculations have demonstrated their predictive power.

*Chern insulators and QAHE*

In spite of the great success in studies of TIs and QSHE, the realization of the QAHE—the quantized version of the anomalous Hall effect (AHE)[69] —had taken a long time. The key to achieving this goal was to find a proper Chern insulator that had a non-zero Chern number. The existence of chiral edge states at the boundary of a Chern insulator can contribute to the dissipationless transport,[21] and result in a kind of QHE without a magnetic field (called QAHE). The Chern insulator and QAHE are fundamentally important because they do not require the protection of TRS, and they are also important in applications because no external magnetic field is needed.

In order to obtain the "twisted" band and a Chern insulator, four necessary conditions must be satisfied simultaneously: (1) 2D system, (2) insulating bulk behavior, (3) ferromagnetic (FM) ordering, and (4) a nonzero Chern number. In real compounds, one or two of these conditions may be easily satisfied, while it is a big challenge to realize them all simultaneously. This is the reason why the existence of a Chern insulator has been theoretically known before the discovery of topological insulators, but its realization had lagged behind.

It is natural to expect that starting from known 2D TIs and then breaking the TRS will be the simplest way to achieve the QAHE. This is because the 2D TIs already satisfy three out of those four conditions, and can be effectively viewed as two Chern insulator layers related by TRS. Breaking of TRS can, in principle, destroy one layer and keep another layer active, which will lead to the QAHE. The theoretical proposal for the possible QAHE in a Mn-doped HgTe quantum well structure[22] is therefore based on this idea. Unfortunately, Mn dopants in HgTe do not order ferromagnetically, thus making the proposal impractical. It is also possible to use the Dirac cone type surface states of a 3D TI, and to construct interfaces between 3D TIs and ferromagnetic insulators.[70] The TRS is broken through the proximity effect, and the gapped Dirac type surface states should contribute to the QAHE. Unfortunately again, this proposal faces problems up until now, such as fine tunning of the interface structure, the strength of the proximity effect, and the well-controlled superlattice structure.

The most realistic proposal to achieve the QAHE is based on magnetic doping in the well-known TI compounds. In 2010, Yu et al. proposed that Cr or Fe doped thin film of $Bi_2Se_3$ or $Bi_2Te_3$ can host the QAHE.[23] Their detailed first-principles calculations showed that: (1) a Cr or Fe dopant has +3 valence and is isovalent to Bi in $Bi_2Se_3$ and $Bi_2Te_3$, and therefore does not introduce additional carriers by substitution; (2) the FM ordering can be established in insulators through the van Vleck susceptibility,[71] which is much enhanced by band inversion; and (3) the resulting exchange splitting is big enough for the realization of the QAHE. This proposal was finally realized experimentally by Chang et al. in 2013 after great efforts,[24] and the predictive power of first-principles calculations is well demonstrated in this example.

Theoretically, there is another promising route to the realization of the QAHE. The QAHE can be achieved from TRS broken Weyl semimetals[72] by forming a quantum well structure.[5] A Weyl semimetal is a 3D system with Fermi surface consists of only the points formed by two non-degenerated bands crossing each other accidentally without any additional symmetry constraints. The low-energy physics around the crossing points is described by Weyl equation. The advantages of this process are that (1) magnetic doping is not necessary; (2) high-temperature QAHE is possible; and (3) QAHE with Chern number larger than one is possible. $HgCr_2Se_4$ has been predicted,[5] based on first-principles calculations, to be a good candidate for a Weyl semimetal, with a Curie temperature of more than 100 K.[73] Its quantum well structure can support the QAHE with a Chern number of 2. Experimental confirmation of this prediction is highly desirable.

### *Large-gap 2D topological insulators*

For 2D $Z_2$ TIs, there are very few known examples, fewer than the number of known 3D TIs[33]. 2D TIs can support the QSHE with helical edge states, whose back-scattering is prohibited by the TRS. Such edge states can serve as a "two-lane highway" for dissipationless electron transport—a phenomena that may be exploited in electronic device applications. The known experimental verifications of the QSHE in HgTe/CdTe[29] and InAs/GaSb[74] quantum-well structures require precisely controlled MBE growth and ultralow temperature, factors that limit further experimental studies and possible applications. In light of these considerations, a "good" 2D TI must meet the following criteria: (1) it must be a good layered material so that it can be easily obtained with natrual cleaving plane; (2) it must have a large 2D bulk bandgap so that its bulk remains insulating even up to room temperature; and (3) it must be chemically stable and its topological properties robust enough against external strains or imperfections.

With these guidelines, several groups have been trying to find appropriate candidate compounds using first-principles calculations. Recently, Weng et al. proposed $ZrTe_5$ and $HfTe_5$.[75] Both of these candidate materials are simple binary compounds with the same layered structure and are known as thermoelectric materials. Weng et al. found that: (1) the interlayer binding energy is relatively weak and comparable to that of graphite; (2) single-layer $ZrTe_5$ and $HfTe_5$ are 2D TI with a bulk direct (indirect) gap as large as 0.4 (0.1) eV; and (3) both compounds are chemically stable, and their topological properties are robust against external distortions.

Other proposed materials include a chemically decorated single-layer honeycomb lattice of Sn/Ge[76,77] and Bi/Sb.[78] Taking the Sn case as an example, the low-buckled Sn honeycomb lattice (Stanene) is predicted to be locally stable by the calculation of its phonon band structure.[76] Similar to silicene,[79] pure stanene is a 2D TI with a bulk bandgap of about 0.1 eV. The bandgap is predicted to be further enhanced up to around 0.3 eV by decorating it with various chemical functional groups. Bandgap larger than 1.0 eV has been recently predicted for yet hypothetical 2D TI BiH and BiF.[78]

### *Correlated topological insulators*

Strong correlation effects can enrich the topological states in condensed matter and lead to "correlated topological states" (CTS), which can be divided into two categories. The first type of CTS can be adiabatically connected to a non-interacting topological state, with their wave functions being single Slater determinants, namely, a matrix form of many electron wave function proposed by J. C. Slater satisfying anti-symmetry requirements and consequently the Pauli principle by changing sign upon exchange of two electrons (or other fermions). This type of CTS is not new, it is a topological state with renormalized quasi-particle band structure. The other type of CTS cannot be adiabatically connected to any non-interacting topological state. The typical examples in this category are fractional Chern insulators,[80-83] fractional TIs, [84,85] quantum spin liquids,[86] and topological Mott Insulators.[87,88] Until now, none of the CTS in the second category has been confirmed in actual materials.

Recently, a special group of mixed valence compounds containing rare-earth elements, including $SmB_6$, $YbB_6$, and $YbB_{12}$, has drawn great attention.[89-91] From the band theory point of view, the "mixed valence" suggests partial occupation of localized $4f$ bands and implies band inversion between them and dispersive $5d$ bands, which may lead to non-trivial topological features. Due to the strong Coulomb interaction among the $4f$ electrons, density functional theory, which has been successfully applied before, cannot be applied directly to these materials. Advanced techniques, which can treat the strong correlation effects more rigorously, are needed. In addition, it is important to understand how to determine the topological nature of a correlated material, which involves many-body wave functions.

The LDA+Gutzwiller method,[92] which combines the LDA with the Gutzwiller method[93-95] (a simplified method for projected many-body wave function) for correlated states, is a powerful tool to treat the CTS. This method starts from the LDA Hamiltonian, and supplements it with the interaction term for the localized $4f$ states, treated using the Gutzwiller-type trial wave functions. The scheme preserves the nice aspect of being variational, however, it is beyond LDA (and also LDA+$U$ [96]) because the total energy now relies on the balance between the renormalized kinetic energy of quasi-particle motion and the local interaction energy, which is configuration-dependent. For details of this method, refer to References 92, 97, and 98. We note that including the complete form of local interactions is essential to obtain the correct electronic structure.

The $Z_2$ classification of band insulators can also be generalized to interacting system by looking at its response to an external electromagnetic field, namely the topological magnetoelectric effect (TME).[99] A correlated insulator is a TI if the θ-angle defined in TME is π. The linear response theory for the coefficient of TME has been developed and simplified by Wang et al.[100] For interacting systems, when the self-energy contains no singularity along the imaginary axis, the formula for the TME coefficient only requires the single particle Green's function at zero frequency, $\hat{g}(k,0)$. Because the singular point for the self-energy along the imaginary axis only appears for a Mott insulator with completely localized $f$-orbitals, which is not the case for Kondo or mixed valence insulators, we can safely apply the above method to them. Therefore, the way to determine the $Z_2$ invariance is by just diagonalizing the Hermite matrix $-\hat{g}(k,0)^{-1} = \hat{H}_0 - \mu_f + \hat{\Sigma}(0)$ and treating the eigenstates with the negative eigenvalues as the "occupied states." In the Gutzwiller approximation, the low-energy single particle Green's function of an interacting system can be expressed by the quasi-particle effective Hamiltonian $H_{eff}$ and the quasi-particle weight $\hat{z}$ (which in general is a matrix) as,

$$\hat{g}(k, i\omega) = \frac{\hat{z}}{i\omega - \hat{H}_{eff} + \mu_f} + \hat{g}_{ic}(k, i\omega) \quad,$$   (4)

where the second term describes the incoherent part of the Green's function, which can be ignored for low frequencies. Therefore, the $Z_2$ invariance in the LDA+Gutzwiller method is determined by the occupied eigenstates of the Gutzwiller effective Hamiltonian $H_{eff}$, which can be interpreted as the band structure of the "quasi-particles." Then the $Z_2$ invariance can be determined by using the same method developed for the non-interacting insulators.[30,52,56]

Using the LDA+Gutzwiller method, we have identified that $SmB_6$ is a topological insulator with the quasi-particle structures strongly renormalized by the Coulomb interaction.[90] The band inversion happens between the itinerant $5d$ and the $4f$  $j = 5/2$ bands at three X points in its Brillouin zone, which leads to three Dirac cones on a generic surface. The topological properties of $SmB_6$ have been verified recently by various experiments, including transport,[101,102] ARPES,[103-105] and STM[106] measurements. Another similar material in this category is $YbB_6$,[91] for which the band inversion occurs between $5d$ and the $4f$  $j = 7/2$ bands. Since the $4f$ shell in $YbB_6$ is nearly fully occupied, the correlation effect as well as the renormalization of the quasi-particle states are much weaker compared to $SmB_6$. Besides the $Z_2$ TIs, the topological crystalline insulator[54,55] can also be found in this class of correlated materials. Applying the LDA+Gutzwiller method again, we found that $YbB_{12}$[91] is such a correlated topological crystalline insulator characterized by a mirror Chern number of 2, which leads to even number of Dirac cones on some specific surfaces.

**Perspectives and conclusion**

There are many other success stories on the selection and design of topological electronic materials, which are not included in this article. As we have mentioned in the introduction, these successes are, on one hand, due to the rapid development in the field of first-principles calculations, and, on the other hand, due to the robustness of topological properties (i.e., its tolerance to error bars). Nevertheless, we are still facing many challenges that require further efforts.

One serious issue is actually an old problem: the local density approximation (LDA) or the generalized gradient approximation (GGA) used in most of the calculations tends to underestimate the bandgap of semiconductors. From the band inversion mechanism, we learn that topological properties crucially depend on the sign of the bandgap, which must be accurately calculated in order to make correct predictions, particularly for narrow gap semiconductors. For example, HgS has been predicted to have an inverted band structure and be a topological insulator based on LDA.[107] This conclusion was challenged recently by accurate GW calculations,[108] which suggested that HgS is a normal semiconductor with a positive bandgap. Therefore, from the calculation point of view, when we encounter a compound with small band inversion, we usually need more advanced and accurate methods, like the GW approximation,[109] the Hybrid functional[110,111] or the modified Becke-Jonhson approximation proposed recently.[112]

Another problem is even more serious: LDA or GGA, which is basically within mean field thory, is qualitatively wrong for strongly correlated systems and should not be used in general for such systems. This is of course a challenging problem and is one of the frontier subjects of modern condensed matter physics. However, as we have discussed, correlated topological insulators are not only fundamentally important, but they are also practically promising from the material point of view; advanced techniques are highly demanded. The computationally cheap and efficient  treatment based on the LDA+$U$ scheme may sometimes improve the situation, but in most of cases, it is still not good enough. We are happy to see progress in recent developments toward this direction, and the newly developed LDA+Gutzwiller and LDA+DMFT (dynamic mean field theory) methods are being increasingly used to treat correlated topological insulators. More efforts in this direction are certainly required.

The great successes of studies on topological insulators are definitely not the end of this field, but rather this should be regarded as an indication of the opening of a much wider field, namely "topological states" or "topological quantum matter." The discovery of topological insulators has stimulated the search for newer quantum states and quantum matter, which are topologically non-trivial. For example, if we consider crystal symmetries of solids, we can get other topological states,[113,114] which may be related to the quantization of polarization.[115,116] The topological semimetal is another interesting state, which is qualitatively different from topological insulators, and has attracted great recent attention.[4–11,42–44] The concept of topology can also be combined with superconductivity, which leads to the "tolopogical superconductors"—a promising and challenging field awaiting extensive exploration.[2,12–15] For this, however, the predictive power of first-principles calculations is rather limited. For all of these exotic topological quantum states, further explorations for better materials are certainly necessary. We are still in the early stages of possible applications of topological quantum states, and materials are the key to advancing the field.

## Acknowledgments

We acknowledge support from the NSF of China and the 973 program of China (No. 2011CBA00108 and 2013CB921700).